\documentclass[12pt]{article}
\usepackage{amssymb}
\usepackage{amsmath}
\usepackage[space]{cite}

\numberwithin{equation}{section}

\newcommand{\be}{\begin{equation}}
\newcommand{\ee}{\end{equation}}
\newcommand{\non}{\nonumber}
\newcommand{\id}{\mathbb{I}}

\newcommand{\tr}{\mathop{\rm tr}\nolimits}
\newcommand{\diag}{\mathop{\rm diag}\nolimits}

\usepackage[usenames,dvipsnames]{color}

\begin{document}

\begin{titlepage}
\strut\hfill UMTG--310
\vspace{.5in}
\begin{center}

{\LARGE Spin chains with boundary inhomogeneities}\\
\vspace{1in}
\large 
Rafael I. Nepomechie \footnote{Physics Department,
P.O. Box 248046, University of Miami, Coral Gables, FL 33124 USA, nepomechie@miami.edu}
and Ana L. Retore  \footnote{School of Mathematics \& Hamilton 
Mathematics Institute, Trinity College Dublin, Dublin, Ireland, retorea@maths.tcd.ie}

\end{center}

\vspace{.5in}

\begin{abstract}
	We investigate the effect of introducing a boundary inhomogeneity 
	in the transfer matrix of an integrable open quantum spin chain.
	We find that it is possible to construct a local Hamiltonian, and to 
	have quantum group symmetry. The boundary inhomogeneity has a 
	profound effect on the Bethe ansatz solution.
\end{abstract}

\end{titlepage}

\setcounter{footnote}{0}

\section{Introduction}\label{sec:intro}

Integrable open spin chains with boundaries have been widely studied
in a variety of contexts, see e.g. \cite{Gaudin:1971zza,
Alcaraz:1987uk, Sklyanin:1988yz, Pasquier:1989kd, Sandow:1994,
Kitanine:2007bi, Zoubos:2010kh, Wang2015} and references therein.
Sklyanin \cite{Sklyanin:1988yz} provided a general recipe for
constructing such models, based on solutions of the bulk
\cite{Jimbo:1989qm} and boundary \cite{Cherednik:1985vs,
Ghoshal:1993tm} Yang-Baxter equations (YBEs), to which we refer here 
as R-matrices and K-matrices, respectively.

It was recently noticed, in the context of the $A^{(1)}_{1}$ R-matrix
(corresponding to a spin chain of XXZ-type) and a trivial K-matrix,
that Sklyanin's construction can be further generalized by introducing
\cite{Nepomechie:2020onv} a boundary inhomogeneity in the transfer
matrix, as in \eqref{topen} below.\footnote{This was actually a side
result of \cite{Nepomechie:2020onv}, which was primarily devoted to
solving $D^{(2)}_{2}$ models.} The corresponding Hamiltonian is
generally not expected to be local; however, by virtue of also having
suitable staggered bulk inhomogeneities, the resulting
Hamiltonian is in fact local.  This model has some further remarkable
features, including quantum group (QG) symmetry \cite{Chari:1994pz}, a
novel Bethe ansatz solution, and a continuum limit described by a
non-compact CFT \cite{Robertson:2020imc}, see also
\cite{Jacobsen:2005xz, Ikhlef:2008zz, Ikhlef:2011ay, Candu:2013fva,
Frahm:2013cma, Bazhanov:2019xvy, Bazhanov:2020dlm} for the
corresponding closed chain.

The goal of the present paper is to explore such models with boundary
inhomogeneities more broadly, particularly by considering higher-rank
R-matrices, as well as non-trivial K-matrices.  For concreteness, we
focus here on the infinite family of $A^{(2)}_{2n}$ R-matrices;
however, we expect that similar results hold for other trigonometric
R-matrices \cite{Bazhanov:1984gu, Bazhanov:1986mu, Jimbo:1985ua} with
crossing symmetry (including $A_{2n-1}^{(2)}\,, B_n^{(1)}\,,
C_n^{(1)}\,, D_n^{(1)}\,, D_{n+1}^{(2)}$).  By introducing suitable
staggered bulk inhomogeneities, we find that the key features of
locality of the Hamiltonian and QG symmetry appearing at rank one can
still be maintained, and again find novel Bethe ansatz solutions.

The outline of this paper is as follows.  In Sec.  \ref{sec:model} we
define the model by constructing its transfer matrix, and we see
explicitly that its Hamiltonian is local.  We briefly discuss the
model's QG symmetry in Sec.  \ref{sec:QG}, and we present its Bethe
ansatz solution in Sec.  \ref{sec:BA}.  We conclude in Sec.
\ref{sec:end} with a brief summary and a list of some interesting
remaining questions.

\section{The model}\label{sec:model}

We begin with a brief review in Sec. \ref{subsec:basic} of the basic 
ingredients that are used in Sec. \ref{subsec:transfer} to construct the transfer matrix.
We derive the corresponding Hamiltonian in Sec. \ref{subsec:hamiltonian}. 

\subsection{Basic ingredients}\label{subsec:basic}

As already noted, the model is constructed from solutions $R(u)$ of the bulk 
YBE \cite{Jimbo:1989qm}
\be
R_{12}(u - v)\,  R_{13}(u)\, R_{23}(v) = R_{23}(v)\, R_{13}(u)\, R_{12}(u - v)
\,,  \label{YBE}
\ee
\noindent
and solutions $K^{R}(u)$ of the corresponding boundary YBE \cite{Cherednik:1985vs,
Ghoshal:1993tm}
\be
R_{12}(u - v)\, K^{R}_1(u)\, R_{21} (u + v)\, K^{R}_2(v)
= K^{R}_2(v)\, R_{12}(u + v)\, K^{R}_1(u)\, R_{21}(u - v)\,,
\label{BYBEm}
\ee
where the notations follow those in \cite{Nepomechie:2018dsn, Nepomechie:2018nvl}.
For concreteness, we take $R(u)$ to be the $A^{(2)}_{2n}$ R-matrices 
($n=1, 2, \ldots$) 
\cite{Bazhanov:1984gu, Bazhanov:1986mu, Jimbo:1985ua}, which for 
$n=1$ was obtained by Izergin and Korepin \cite{Izergin:1980pe}; 
we use the specific form of these R-matrices given in Appendix A of 
\cite{Nepomechie:2018dsn}, with anisotropy parameter $\eta$. These 
R-matrices have the following additional important properties: periodicity
\be
R(u+2 i \pi) = R(u) \,;
\label{periodicity}
\ee
unitarity
\be
R_{12}(u)\ R_{21}(-u) = \xi(u)\, \xi(-u)\, \id\otimes\id  \,, \qquad
\xi(u) = 2\sinh(\frac{u}{2} -2\eta) \cosh(\frac{u}{2}-(2n+1)\eta)\,;
\label{unitarity}
\ee
regularity
\be
R(0) = \xi(0)\, {\cal P} \,,
\label{regularity}
\ee
where ${\cal P}$ is the permutation matrix; $PT$ symmetry
\be
R_{21}(u) \equiv {\cal P}_{12}\, R_{12}(u)\, {\cal P}_{12} 
= R_{12}^{t_1 t_2}(u) \,;
\label{PT}
\ee
and crossing symmetry
\be
R_{12}(u)=V_1\, R_{12}^{t_2}(-u-\rho)\, V_1
= V_2^{t_2}\, R_{12}^{t_1}(-u-\rho)\, V_2^{t_2} \,, \qquad 
\rho= -2(2n+1) \eta  - i \pi \,,
\label{crossing}
\ee
where the matrix $V$ is given by
\begin{equation}
V=\sum_{\alpha=1}^{2n+1}e^{(\bar{\alpha}-\bar{\alpha}^\prime)\eta}e_{\alpha,\alpha^{\prime}}, 
\quad \alpha^\prime=2n+2-\alpha,\quad  \alpha=1,..., 2n+1 \,,
\end{equation}
and
\begin{align}
\bar{\alpha} &=\left\{ \begin{array}{ll}
\alpha+\frac{1}{2} & 1\le\alpha<n+1\\
\alpha & \alpha=n+1\\
\alpha-\frac{1}{2} & n+1<\alpha\le 2n+1
\end{array} \right..
\label{eq:bar}		     
\end{align}

We take the right K-matrices to be the diagonal matrices \cite{Batchelor:1996np, 
LimaSantos:2002ui, Malara:2004bi}
\begin{align}
K^{R}(u) &= \diag \big( \underbrace{e^{-u}\,, \ldots\,, 
	e^{-u}}_{p}\,, \underbrace{\frac{\gamma e^{u} + 1}{\gamma  + 
		e^{u}}\,, \ldots\,, \frac{\gamma e^{u} + 1}{\gamma  + 
		e^{u}}}_{2n+1-2p}\,, \underbrace{e^{u}\,, \ldots\,, 
	e^{u}}_{p}\big) \,,  \label{KRa}\\
	 & \gamma = \gamma_{0}\, 
	 e^{(4p+2)\eta + \frac{1}{2}\rho}\,, \qquad p = 0, 1, \ldots, n\,,  
	 \qquad \gamma_{0} = \pm 1 \,,
\label{KRb}
\end{align}
which for $n=1$ was obtained in \cite{Mezincescu:1990ui}. We emphasize that 
these K-matrices depend on two boundary parameters $p$ and 
$\gamma_{0}$, which can take the set of discrete values noted in \eqref{KRb}.
Moreover, for the left K-matrices, we take \cite{Nepomechie:2018dsn, Mezincescu:1990ui}
\be
K^{L}(u) = K^{R}(-u-\rho)\, M\,, \qquad M = V^{t}\, V \,,
\ee 
which corresponds to imposing the ``same'' boundary conditions on the two ends. 

For later reference, we note here the useful identity \cite{Ahmed:2017mqq}
\be
\tr_{0} K^{L}_{0}(u)\, R_{01}(2u)\, {\cal P}_{01} = f(u)\, V_{1}\, 
K^{R}_{1}(u)\, V_{1}\,,
\label{trid}
\ee
with
\begin{align}
f(u) &= - 4 
\sinh(\tfrac{u}{2}-\tfrac{1}{2}(2n-1)\eta - \gamma_{0} \tfrac{i \pi}{4})
\sinh(\tfrac{u}{2}-\tfrac{1}{2}(2n+3)\eta + \gamma_{0} \tfrac{i 
\pi}{4}) \non \\
&\quad \times \sinh(u-(4n+2)\eta)\, \frac{\sinh(\tfrac{u}{2}+\tfrac{1}{2}(2n-4p-1)\eta - \gamma_{0} \tfrac{i \pi}{4})}
{\sinh(\tfrac{u}{2}-\tfrac{1}{2}(6n-4p+1)\eta - \gamma_{0} \tfrac{i 
\pi}{4})} \,.
\label{funcf}
\end{align}

\subsection{Transfer matrix}\label{subsec:transfer}

We consider the following open-chain transfer matrix for a spin chain 
of length $N$ \cite{Nepomechie:2020onv}
\be
t(u; \{\theta_{l}\}, u_{0}) = \tr_{0} \Big\{  \bar{K}^{L}_{0}(u)\, T_{0}(u; \{\theta_{l}\})\, 
\bar{K}^{R}_{0}(u)\, \widehat{T}_{0}(u+u_{0}; \{\theta_{l}\})\Big\}\,,
\label{topen}
\ee
whose key difference with respect to the transfer matrix in
\cite{Sklyanin:1988yz} is the shift by $u_{0}$ in the argument of
$\widehat{T}$, which can be regarded as a boundary inhomogeneity.  We
shall see that this seemingly minor change in the transfer matrix in
fact has a profound impact on the model.  The monodromy matrices are
given as usual by
\begin{align}
T_{0}(u; \{\theta_{l}\}) &= R_{0 N}(u-\theta_{N})\, \ldots R_{0 
1}(u-\theta_{1}) \,, \non\\
\widehat{T}_{0}(u; \{\theta_{l}\}) &= R_{1 0}(u+\theta_{1})\, \ldots 
R_{N 0}(u+\theta_{N}) \,,
\label{monodromyTinhom2}
\end{align}
where $\{\theta_{l}\}$ are bulk inhomogeneities. The right K-matrix 
$\bar{K}^{R}_{0}(u)$ in \eqref{topen} satisfies a generalized boundary YBE
\be
R_{12}(u-v)\, \bar{K}^{R}_{1}(u)\, R_{21}(u+v+u_{0})\, 
\bar{K}^{R}_{2}(v) = 
\bar{K}^{R}_{2}(v)\, R_{12}(u+v+u_{0})\, \bar{K}^{R}_{1}(u)\, 
R_{21}(u-v) \,,
\label{BYBEm-gen}
\ee
which, compared with \eqref{BYBEm}, has a shift by $u_{0}$ in the R-matrix whose argument has the 
sum of rapidities. The generalized boundary YBE \eqref{BYBEm-gen} can be 
mapped to the standard one \eqref{BYBEm} by performing the shifts $u \mapsto u - u_{0}/2$ 
and $v \mapsto v - u_{0}/2$, and identifying 
$\bar{K}^{R}(u- u_{0}/2) = K^{R}(u)$. Hence, we set
\be
\bar{K}^{R}(u) = K^{R}(u+ \tfrac{u_{0}}{2}) \,,
\ee
with $K^{R}(u)$ given by \eqref{KRa}. Setting \cite{Nepomechie:2020onv} 
\be
\bar{K}^{L}(u) = \bar{K}^{R}(-u-\rho-u_{0})\, M\,,
\ee
the transfer matrix \eqref{topen} can be shown to satisfy the commutativity property
\be
\left[t(u; \{\theta_{l}\}, u_{0}) \,, t(v; \{\theta_{l}\}, u_{0}) \right] = 0 \,,
\label{commutativity}
\ee
which is the hallmark of quantum integrability.

In terms of the $\bar{K}$-matrices, the identity (\ref{trid}) reads
\be
\tr_{0} \bar{K}^{L}_{0}(u)\, R_{01}(2u+u_{0})\, {\cal P}_{01} = f(u+ \tfrac{u_{0}}{2})\, V_{1}\, 
\bar{K}^{R}_{1}(u)\, V_{1}\,,
\label{trid-gen}
\ee
where $f(u)$ is given by \eqref{funcf}.

An important observation is that the presence of a boundary 
inhomogeneity affects the crossing relation of the transfer matrix. 
Indeed, the crossing relation now becomes
\be
t(-u-\rho-u_{0}; \{\theta_{l}\}, u_{0}) = t(u; \{\theta_{l}\}, u_{0}) \,,
\label{tcrossing}
\ee
i.e. there is an additional $u_{0}$-dependent shift.

For generic values of boundary and bulk inhomogeneities, the transfer
matrix \eqref{topen} does not generate a local Hamiltonian (i.e.,
whose range of interactions is independent of $N$).  Following
\cite{Nepomechie:2020onv}, we henceforth set these inhomogeneities to
\be
u_{0} = i \pi\,, \qquad \theta_{l} = \begin{cases}
-i \pi & \text{for $l=$ odd}\\
0 & \text{for $l=$ even}
\end{cases} \,.
\label{inhomgen}
\ee
Note that the boundary inhomogeneity $u_{0}$ is the half-period of the
R-matrix \eqref{periodicity}, and the bulk inhomogeneities are
staggered \footnote{Staggered models date back at least to \cite{Baxter:1982b}.}.  
The same transfer matrix but with no bulk or boundary
inhomogeneities $(u_{0} = \theta_{l} = 0)$, to which we refer as the
``homogeneous case'', was investigated in \cite{Nepomechie:2018dsn,
Nepomechie:2018nvl}.

To summarize, we consider the transfer matrix \eqref{topen} with 
inhomogeneity values given by \eqref{inhomgen}; it depends on the 
discrete parameters $N \in \{1, 2, \ldots\}$,
$n \in \{1, 2, \ldots\}$, $p \in \{0, 1, \ldots, n\}$ and
$\gamma_{0} \in \{-1, +1\}$, as well as the continuous parameters 
$u$ and $\eta$ .

\subsection{Hamiltonian}\label{subsec:hamiltonian}

We define the $N$-site Hamiltonian by\footnote{We henceforth suppress displaying the 
dependence of the transfer matrix on the inhomogeneities, which are 
given by \eqref{inhomgen}.}
{\allowdisplaybreaks 
\be
{\cal H}^{(N)} = \frac{d}{du}\log\left(t(u)\right)\Big\vert_{u=0}
= t^{-1}(0) \frac{d}{du}t(u)\Big\vert_{u=0} \,.
\label{Hdef}
\ee
Note that this is the usual recipe for a closed-chain, rather than an open-chain, Hamiltonian.
(For an open chain, usually $t(0) \propto \id$ \cite{Sklyanin:1988yz}, hence the definition \eqref{Hdef} reduces to 
$t'(0)$. However, here $t(0)$ is not proportional to $\id$, thus 
these two definitions are not equivalent; and the latter definition does not yield a local Hamiltonian.)
Using \eqref{trid-gen} and the identity 
\be
t(0)\, t(i\pi) = \xi^{2N}(0)\, \xi^{2N-2}(i\pi)\, 
f(\tfrac{i\pi}{2})\, f(\tfrac{3i\pi}{2})\, \id \,,
\ee
we obtain, after a long calculation, the following Hamiltonian for even 
values of $N>2$ 

\begin{align}
{\cal H}^{(N=\text{even})} &= \frac{1}{\xi(0)\, \xi^{4}(i\pi)}\, \bar{K}^{R}_{1}(i\pi)\, R_{32}\, 
R_{31}\, h_{12}\, R_{13}\, R_{23}\, \bar{K}^{R}_{1}(0) \non \\
&\qquad + \frac{1}{\xi(0)}\, V_{N}\, 
\bar{K}^{R}_{N}(i\pi)\, V_{N}\, h_{N-1,N}\, V_{N}\, 
\bar{K}^{R}_{N}(0)\, V_{N} \non \\
&\qquad + \frac{1}{\xi(0)\, \xi^{2}(i\pi)}\, R_{N-1,N-2}\, h_{N-2,N}\, R_{N-2,N-1} \non \\
&\qquad + \frac{1}{\xi(0)\, \xi^{2}(i\pi)}\, \sum_{j=1,3,\ldots}^{N-3}R_{j+2,j+1}\, h_{j,j+2}\, R_{j+1,j+2} \non \\
&\qquad + \frac{1}{\xi(0)\, \xi^{6}(i\pi)}\, \sum_{j=2,4,\ldots}^{N-4}R_{j+3,j+2}\, R_{j+1,j}\, R_{j+3,j}\,
h_{j,j+2}\, R_{j,j+3}\, R_{j,j+1}\, R_{j+2,j+3}\,\non \\
&\qquad + \frac{1}{\xi^{2}(i\pi)}\, \sum_{j=2,4,\ldots}^{N-2} \bar{h}_{j,j+1} \non \\
&\qquad + \frac{1}{\xi^{4}(i\pi)}\, \bar{K}^{R}_{1}(i\pi)\, R_{32}\, \bar{h}_{13}\, R_{23}\, 
\bar{K}^{R}_{1}(0) \non \\
&\qquad + \frac{1}{\xi^{6}(i\pi)}\,\sum_{j=2,4,\ldots}^{N-4}R_{j+3,j+2}\, R_{j+1,j}\, 
\bar{h}_{j,j+3}\, R_{j,j+1}\, R_{j+2,j+3}\,\non \\
&\qquad + \bar{K}^{R}_{1}(i\pi)\, \bar{K}^{R'}_{1}(0)
+ V_{N}\, \bar{K}^{R}_{N}(i\pi)\, \bar{K}^{R'}_{N}(0)\, V_{N} \non \\
&\qquad + \frac{f'(\tfrac{i\pi}{2})}{f(\tfrac{i\pi}{2})}\, \,\id \,,
\label{Heven}
\end{align}}
where we have introduced the following short-hand notations
\be
h_{ij} = {\cal P}_{ij}\, R'_{ij}(0)\,, \qquad \bar{h}_{ij} = 
R_{ji}(i\pi)\, R'_{ij}(i\pi)\,, \qquad R_{ij} = R_{ij}(i\pi)\,,
\ee
and a prime denotes differentiation with respect to the spectral
parameter $u$.  Note that the range of interactions in this Hamiltonian
does not exceed 4 sites. For the case $N=2$, we obtain
\begin{align}
{\cal H}^{(N=2)} &= \frac{1}{\xi(0)}\, \bar{K}^{R}_{1}(i\pi)\, h_{12}\, \bar{K}^{R}_{1}(0) 
+ \frac{1}{\xi(0)}\, V_{2}\, 
\bar{K}^{R}_{2}(i\pi)\, V_{2}\, h_{12}\, V_{2}\, \bar{K}^{R}_{2}(0)\, V_{2} \non \\
&\qquad + \bar{K}^{R}_{1}(i\pi)\, \bar{K}^{R'}_{1}(0)
+ V_{2}\, \bar{K}^{R}_{2}(i\pi)\, \bar{K}^{R'}_{2}(0)\, V_{2} 
+ \frac{f'(\tfrac{i\pi}{2})}{f(\tfrac{i\pi}{2})}\, \,\id \,.
\label{2}
\end{align}

A similar computation for odd values of $N>1$ gives
{\allowdisplaybreaks 
\begin{align}
{\cal H}^{(N=\text{odd})} &= \frac{1}{\xi(0)\, \xi^{4}(i\pi)}\, \bar{K}^{R}_{1}(i\pi)\, R_{32}\, 
R_{31}\, h_{12}\, R_{13}\, R_{23}\, \bar{K}^{R}_{1}(0) \non \\
&\qquad + \frac{1}{\xi(0)\, \xi^{2}(i\pi)}\, V_{N}\, 
\bar{K}^{R}_{N}(0)\, V_{N}\, R_{N,N-1}\, h_{N,N-1}\, R_{N-1,N}\, V_{N}\, 
\bar{K}^{R}_{N}(i\pi)\, V_{N} \non \\
&\qquad + \frac{1}{\xi(0)\, \xi^{2}(i\pi)}\, 
\sum_{j=1,3,\ldots}^{N-2}R_{j+2,j+1}\, h_{j,j+2}\, R_{j+1,j+2} \non \\
&\qquad + \frac{1}{\xi(0)\, \xi^{6}(i\pi)}\, \sum_{j=2,4,\ldots}^{N-3}R_{j+3,j+2}\, R_{j+1,j}\, R_{j+3,j}\,
h_{j,j+2}\, R_{j,j+3}\, R_{j,j+1}\, R_{j+2,j+3}\,\non \\
&\qquad + \frac{1}{\xi^{2}(i\pi)}\, \sum_{j=2,4,\ldots}^{N-3} 
\bar{h}_{j,j+1} \non \\ 
&\qquad + \frac{1}{\xi^{4}(i\pi)}\, \bar{K}^{R}_{1}(i\pi)\, R_{32}\, \bar{h}_{13}\, R_{23}\, 
\bar{K}^{R}_{1}(0) \non \\ 
&\qquad + \frac{1}{\xi^{6}(i\pi)}\,\sum_{j=2,4,\ldots}^{N-3}R_{j+3,j+2}\, R_{j+1,j}\, 
\bar{h}_{j,j+3}\, R_{j,j+1}\, R_{j+2,j+3}\,\non \\
&\qquad + \bar{K}^{R}_{1}(i\pi)\, \bar{K}^{R'}_{1}(0)
+ V_{N}\, \bar{K}^{R}_{N}(i\pi)\, \bar{K}^{R'}_{N}(0)\, V_{N} 
\non \\
&\qquad + \frac{f'(\tfrac{i\pi}{2})}{f(\tfrac{i\pi}{2})}\, \,\id 
\,.
\label{Hodd}
\end{align}}
The range of interactions again does not exceed 4 sites. We conclude 
that the Hamiltonian is local.

\section{Quantum group symmetry}\label{sec:QG}

The transfer matrix \eqref{topen} with inhomogeneities
\eqref{inhomgen} has the QG symmetry $U_{q}(B_{n-p}) \otimes
U_{q}(C_{p}) $, corresponding to removing the $p^{th}$ node from the
$A^{(2)}_{2n}$ Dynkin diagram, as follows from arguments similar to
those for the homogeneous case 
\cite{Nepomechie:2018dsn}.\footnote{The 
gauge transformations for the K-matrices are now given by
$$ \tilde{\bar{K}}^R(u,p) = B(u + \tfrac{u_0}{2},p)\, \bar{K}^R(u,p)\, 
B(u + \tfrac{u_0}{2},p)\,, \qquad
\tilde{\bar{K}}^L(u,p) = B(-u - \tfrac{u_0}{2},p)\, \bar{K}^L(u,p)\, 
B(-u -\tfrac{u_0}{2},p)\,, $$
where $B(u,p)$ is given by Eq. (3.3) in \cite{Nepomechie:2018dsn}.}
The
``left'' algebra $B_{n-p}$ (with $p=0, 1, \ldots, n-1$) has generators
\begin{align}
H^{(l)}_{j}(p) &= e_{p+j,p+j} - e_{2n+2-p-j,2n+2-p-j}\,, \non\\
E^{+\, (l)}_{j}(p) &= e_{p+j,p+j+1} + e_{2n+1-p-j,2n+2-p-j}\,, \non\\
E^{-\, (l)}_{j}(p) &= \left(E^{+\, (l)}_{j}(p)\right)^{t}\,, \qquad j = 1, \ldots, n-p\,,  
\end{align}
and the ``right'' algebra $C_{p}$ (with $p=1, 2, \ldots, n$) has generators
\begin{align}
H^{(r)}_{j}(p) &= -e_{p+1-j,p+1-j} + e_{2n+1-p+j,2n+1-p+j}\,, \non\\
E^{+\, (r)}_{j}(p) &= 
\begin{cases}
e_{p-j,p+1-j} + e_{2n+1-p+j,2n+2-p+j} & \text{for }  1\le j \le p-1\\
\sqrt{2} e_{2n+1,1} & \text{for } j=p
\end{cases}	\,, \non\\ 
E^{-\, (r)}_{j}(p) &= \left( E^{+\, (r)}_{j}(p) \right)^{t} \,, \qquad j = 1, \ldots, p \,, 
\end{align}
where $e_{ij}$ are the elementary $(2n+1) \times (2n+1)$ matrices 
with elements $(e_{i j})_{\alpha \beta} = \delta_{i, \alpha} \delta_{j, \beta}$. 
The coproducts for the ``left'' generators are given by
\begin{align}
\Delta(H^{(l)}_{j}) &= H^{(l)}_{j} \otimes \id + \id \otimes H^{(l)}_{j} \,, \qquad j = 
1, \ldots,  n-p \,, \non \\
\Delta(E^{\pm\, (l)}_{j}) &= E^{\pm\, (l)}_{j} \otimes e^{(\eta + i \pi)
H^{(l)}_{j} - \eta H^{(l)}_{j+1}} + e^{-(\eta + i \pi)
H^{(l)}_{j} + \eta H^{(l)}_{j+1}} \otimes E^{\pm\, (l)}_{j} \,,
\qquad j = 1, \ldots, n-p-1 \,, \non  \\
\Delta(E^{\pm\, (l)}_{n-p}) &= 
E^{\pm\, (l)}_{n-p} \otimes e^{(\eta + i \pi) 
H^{(l)}_{n-p}} + e^{-(\eta + i \pi) 
H^{(l)}_{n-p} } \otimes E^{\pm\, (l)}_{n-p} 
\,,
\label{coproductleft}
\end{align}
and the coproducts for the ``right'' generators are given by
\begin{align}
\Delta(H^{(r)}_{j}) &= H^{(r)}_{j} \otimes \id + \id \otimes H^{(r)}_{j} \,, \qquad j = 
1, \ldots,  p \,, \non \\
\Delta(E^{\pm\, (r)}_{j}) &= E^{\pm\, (r)}_{j} \otimes e^{(\eta + i \pi)
H^{(r)}_{j} - \eta H^{(r)}_{j+1}} + e^{-(\eta + i \pi)
H^{(r)}_{j} + \eta H^{(r)}_{j+1}} \otimes E^{\pm\, (r)}_{j} \,, \qquad j 
= 1, \ldots, p-1 \,, \non  \\
\Delta(E^{\pm\, (r)}_{p}) &=  E^{\pm\, (r)}_{p} \otimes e^{2 \eta 
H^{(r)}_{p}}  - \, e^{-2 \eta 
H^{(r)}_{p}} \otimes E^{\pm\, (r)}_{p}  \,.
\label{coproductright}
\end{align}
These expressions for the coproducts are the same as in
\cite{Nepomechie:2018dsn} (where many further details can also be found), except
for the relative minus sign in $\Delta(E^{\pm\, (r)}_{p})$
\eqref{coproductright}.

Due to the relative minus sign in 
$\Delta(E^{\pm\, (r)}_{p})$ \eqref{coproductright}, this coproduct does not obey the 
standard co-associativity property. Indeed,
\begin{align}
	(\id \otimes \Delta) \Delta(E^{\pm\, (r)}_{p}) &= (\id \otimes 
	\Delta)\left( E^{\pm\, (r)}_{p} \otimes e^{2 \eta H^{(r)}_{p}}  - 
	e^{-2 \eta H^{(r)}_{p}} \otimes E^{\pm\, (r)}_{p} \right) \non \\
	&= E^{\pm\, (r)}_{p} \otimes e^{2\eta \Delta(H^{(r)}_{p})} - 
	e^{-2\eta H^{(r)}_{p}} \otimes \Delta(E^{\pm\, (r)}_{p}) \non \\
	&= \Delta(E^{\pm\, (r)}_{p}) \otimes e^{2\eta H^{(r)}_{p}} 
	+ e^{-2\eta \Delta(H^{(r)}_{p})} \otimes E^{\pm\, (r)}_{p} \non \\
	&\ne (\Delta\otimes \id) \Delta(E^{\pm\, (r)}_{p}) \,, 
\end{align}
which suggests that there is instead an underlying quasi-Hopf algebra 
structure \cite{Chari:1994pz, Gainutdinov:2021}.
We define the higher coproducts for $E^{\pm\, (r)}_{p}$ recursively by
\begin{align}
\Delta_{N}(E^{\pm\, (r)}_{p}) &= (\id \otimes \Delta) 
\Delta_{N-1}(E^{\pm\, (r)}_{p}) \non\\
&= E^{\pm\, (r)}_{p} \otimes e^{2 \eta 
\Delta_{N-1}(H^{(r)}_{p})}  - \, e^{-2 \eta 
H^{(r)}_{p}} \otimes \Delta_{N-1}(E^{\pm\, (r)}_{p})\,, \qquad N>2\,,
\end{align}	
where $\Delta_{2} = \Delta$.
	
The $N$-fold 
coproducts of the ``left'' and ``right'' generators commute with the transfer 
matrix \eqref{topen}, \eqref{inhomgen}
\begin{align}
	\left[ \Delta_{N}(H^{(l)}_{j}) \,, t(u) \right] &= \left[ 
	\Delta_{N}(E^{\pm\, (l)}_{j}) \,, t(u) \right] = 0 \,, \qquad j=1, 
	\ldots, n-p\,, \qquad p =0,\ldots, n-1 \,, \non \\
	\left[ \Delta_{N}(H^{(r)}_{j}) \,, t(u) \right] &= \left[ 
	\Delta_{N}(E^{\pm\, (r)}_{j}) \,, t(u) \right] = 0 \,, \qquad j=1, 
	\ldots, p \,, \qquad p =1, \ldots, n\,,
	\label{QGt}
\end{align}
for both values $\gamma_{0}=\pm 1$.
At least for real values of the anisotropy parameter
$\eta$, the rich degeneracies in the spectrum of the transfer matrix 
are completely accounted for by its QG symmetry, as in the homogeneous 
case \cite{Nepomechie:2018dsn}.

\section{Analytical Bethe ansatz}\label{sec:BA}

The eigenvalues of the transfer matrix \eqref{topen}, 
\eqref{inhomgen} can be determined by analytical Bethe ansatz \cite{Reshetikhin:1987}
similarly to the homogeneous case \cite{Nepomechie:2018nvl}; however, 
there are some surprises. Indeed, let $|\Lambda^{(m_{1}, \ldots, 
m_{n})}\rangle$ be simultaneous eigenvectors of the transfer matrix 
and Cartan generators
\begin{align}
t(u)\, |\Lambda^{(m_{1}, \ldots\,, m_{n})}\rangle &= \Lambda^{(m_{1}, 
\ldots, m_{n})}(u)\,
|\Lambda^{(m_{1}, \ldots, m_{n})}\rangle \,, \non \\
\Delta_{N}(H_{i}^{(l)}(p))\, |\Lambda^{(m_{1}, \ldots, m_{n})}\rangle &= 
h^{(l)}_{i}\,
|\Lambda^{(m_{1}, \ldots, m_{n})}\rangle \,, \qquad i = 1, \ldots, 
n-p \,, \non \\
\Delta_{N}(H_{i}^{(r)}(p))\, |\Lambda^{(m_{1}, \ldots, m_{n})}\rangle &= 
h^{(r)}_{i}\,
|\Lambda^{(m_{1}, \ldots, m_{n})}\rangle \,, \qquad i = 1, \ldots, p 
\,.
\label{eigenvalueproblem}
\end{align}

We propose that the eigenvalues of the transfer matrix 
for general values of $n$, $p$ and $\gamma_{0}$ are given by the 
following TQ-equation
\begin{align}
\Lambda^{(m_1,...,m_n)}(u)&=\phi(u,p) \Bigg\{ A(u)\, z_0(u)\, 
y_0(u,p)\, 
\left[-\sinh(u-4\eta) \sinh(u-2(2n+1)\eta)\right]^{N} \non\\
&\qquad +\tilde{A}(u)\, \tilde{z}_0(u)\, \tilde{y}_0(u,p)\, 
\left[-\sinh(u) \sinh(u-2(2n-1)\eta)\right]^{N}\non\\
&\qquad +\Big\{\sum_{l=1}^{n-1}\left[z_l(u)\, y_l(u,p)\, B_l(u)+\tilde{z}_l(u)\, 
\tilde{y}_l(u,p)\, \tilde{B}_l(u)\right] \non\\
&\qquad+ w(u)\, y_n(u,p)\, B_n(u) 
\Big\}\, \left[-\sinh(u) \sinh(u-2(2n+1)\eta)\right]^{N} \Bigg\}\,.
\label{eigenvalue}
\end{align}
The overall factor $\phi(u,p)$ is given by 
\be
\phi(u,p) = 
 \left(\frac{\gamma e^{u+\frac{i \pi}{2}} + 1}{\gamma + e^{u+\frac{i \pi}{2}}}\right)\left(\frac{\gamma 
e^{-u-\rho-\frac{i \pi}{2}} + 1}{\gamma + e^{-u-\rho-\frac{i \pi}{2}}}\right)\,,
\ee
where $\gamma$ is defined in \eqref{KRb}. The tilde denotes crossing 
e.g. $\tilde{A}(u)=A(-u-\rho-i\pi)$, in view of the crossing relation \eqref{tcrossing}.
The functions $A(u)$ and $B_l(u)$ are defined as
\begin{align}
A(u) &=\frac{Q^{[1]}(u+2\eta)}{Q^{[1]}(u-2\eta)}\,, \non \\
B_l(u) 
&=\frac{Q^{[l]}(u-2(l+2)\eta)}{Q^{[l]}(u-2l\eta)}\frac{Q^{[l+1]}(u-2(l-1)\eta)}{ Q^{[l+1]}(u-2(l+1)\eta)}\,, \qquad
l=1,...,n-1 \,, \non \\
B_{n}(u) 
&=\frac{Q^{[n]}(u-2(n+2)\eta)}{Q^{[n]}(u-2n\eta)}\frac{Q^{[n]}(u-2(n-1)\eta+i \pi)}{Q^{[n]}(u-2(n+1)\eta+i\pi)} \,,
\label{AB}
\end{align}
where the functions $Q^{[l]}(u)$ are given by
\begin{align}
Q^{[l]}(u)&=\prod_{j=1}^{m_l}\sinh\left(\tfrac{1}{2}(u-u_j^{[l]})\right)\cosh\left(\tfrac{1}{2}(u+u_j^{[l]})\right)\,,
\qquad  Q^{[l]}(-u) = Q^{[l]}(u+ i \pi) \,, \non \\
&\qquad\qquad l = 1, \ldots, n \,,
\label{Qfunc}
\end{align}
whose zeros $\{ u_{j}^{[l]} \}$  remain to be determined.
The functions $z_l(u)$ and $w(u)$ are given by 
\begin{align}
z_l(u) &=\frac{\cosh(u)\cosh(u-2(2n+1)\eta)\sinh(u-(2n-1)\eta)}{\cosh(u-2l\eta)\cosh(u-2(l+1)\eta)
\sinh(u-(2n+1)\eta)}\,, \non\\
w(u) &=\frac{\cosh(u)\cosh(u-2(2n+1)\eta)}{\cosh(u-2n\eta) 
\cosh(u-2(n+1)\eta)} \,,
\label{zw}
\end{align}
and the functions $y_l(u,p)$ are given by 
\be
y_l(u,p) =\begin{cases}
F(u) \quad \mbox{for} \quad 0\le l \le p-1\\
G(u) \quad \mbox{for} \quad p\le l \le n
\end{cases}\,,
\label{yl}
\ee
where
\begin{align} 
G(u) &=\frac{\cosh\left(\frac{1}{2}(u-(2n-1)\eta + i \pi \varepsilon)\right)
\cosh\left(\frac{1}{2}(u-(2n+3)\eta + i \pi \varepsilon)\right)}
{\cosh\left(\frac{1}{2}(u-(2n-4p-1)\eta + i \pi \varepsilon)\right) 
\cosh\left(\frac{1}{2}(u-(2n+4p+3)\eta + i \pi \varepsilon)\right)}\,, \non \\
F(u) &= -\left(\frac{\sinh\left(\frac{1}{2}(u+(2n-4p-1)\eta + i \pi \varepsilon)\right)}
{\cosh\left(\frac{1}{2}(u-(2n-1)\eta + i \pi \varepsilon)\right)}\right)^{2}\, G(u) \,,
\label{GF}
\end{align}
with 
\be
\varepsilon=\frac{1}{2}(1-\gamma_{0}) \in \{0, 1 \} \,.
\label{varepsilon}
\ee
The Bethe equations for the zeros $\{ u_{k}^{[l]} \}$ of the 
Q-functions, which we determine by requiring that
the transfer-matrix eigenvalues (\ref{eigenvalue}) have vanishing
residues at the poles $u=u_{k}^{[l]}+2 l \eta$, are given by
\be
\left[\frac{\sinh\left(u_k^{[1]}+2\eta\right)}
{\sinh\left(u_k^{[1]}-2\eta\right)}\right]^{N}\Phi_{1,p,n}(u_k^{[1]}) =
\frac{Q_k^{[1]}\left(u_k^{[1]}+4\eta\right)}{Q_k^{[1]}\left(u_k^{[1]}-4\eta\right)}
\frac{Q^{[2]}\left(u_k^{[1]}-2\eta\right)}
{Q^{[2]}\left(u_k^{[1]}+2\eta\right)}\,, \quad k = 1, \ldots, m_{1}\,, \label{BE1}
\ee 
\begin{align}
\Phi_{l,p,n}(u_k^{[l]}) &=\frac{Q^{[l-1]}\left(u_k^{[l]}-2\eta\right)}{Q^{[l-1]}
\left(u_k^{[l]}+2\eta\right)}\frac{Q_k^{[l]}\left(u_k^{[l]}+4\eta\right)}
{Q_k^{[l]}\left(u_k^{[l]}-4\eta\right)}
\frac{Q^{[l+1]}\left(u_k^{[l]}-2\eta\right)}{Q^{[l+1]}\left(u_k^{[l]}+2\eta\right)}\,, 
\quad k = 1, \ldots, m_{l}\,,   \non \\
& \qquad\qquad\qquad l=2,...,n-1 \,, \label{BE2} \\
\Phi_{n,p,n}(u_k^{[n]}) &=\frac{Q^{[n-1]}\left(u_k^{[n]}-2\eta\right)}
{Q^{[n-1]}\left(u_k^{[n]}+2\eta\right)}\frac{Q_k^{[n]}\left(u_k^{[n]}+4\eta\right)}
{Q_k^{[n]}\left(u_k^{[n]}-4\eta\right)}
\frac{Q_k^{[n]}\left(u_k^{[n]}-2\eta+i\pi\right)}
{Q_k^{[n]}\left(u_k^{[n]}+2\eta+i\pi\right)}\,, 
\quad k = 1, \ldots, m_{n}\,,\label{BE3}
\end{align}
where $Q^{[l]}(u)$ is given by (\ref{Qfunc}), and $Q_k^{[l]}(u)$ is 
defined by a similar product with the $k^{th}$ term omitted
\be
Q_k^{[l]}(u)=\prod_{j=1, j \ne 
k}^{m_l}\sinh\left(\tfrac{1}{2}(u-u_j^{[l]})\right)\cosh\left(\tfrac{1}{2}(u+u_j^{[l]})\right)\,.
\label{Qk}
\ee
Finally, the important factor $\Phi_{l,p,n}(u)$ appearing in the Bethe 
equations is given by
\begin{align}
\Phi_{l,p,n}(u) &=\frac{y_{l}(u + 2 l \eta,p)}{y_{l-1}(u + 2 l \eta,p)} = 
\begin{cases}
    \frac{G(u + 2 p \eta)}{F(u + 2 p \eta)} & \text{for }  l=p \\
    1 & \text{for }  l\ne p
\end{cases} \,, \non \\
&= \left(\frac{\sinh\left(\tfrac{1}{2}(u-\delta_{l,p}[(2n-2p-1)\eta 
+ i \pi \delta_{\varepsilon,0}])\right)}
{\sinh\left(\tfrac{1}{2}(u+\delta_{l,p}[(2n-2p-1)\eta 
+ i \pi \delta_{\varepsilon,1}] )\right)} \right)^{2} \,,
\end{align}
where $\varepsilon$ is defined in \eqref{varepsilon}.
Note that $\Phi_{l,p,n}(u)$ is different from 1 only if $l=p$.

The energy is given, in view of \eqref{Hdef} and \eqref{eigenvalue}, by
\begin{align}
	E &= \frac{d}{du}\log\left(\Lambda^{(m_{1}, 
\ldots, m_{n})}(u)\right)\Big\vert_{u=0} \non \\
&=-\sum_{j=1}^{m_{1}}\frac{\sinh(4\eta)}
{\sinh(u_j^{[1]}+2\eta)\sinh(u_j^{[1]}-2\eta)}
- \frac{N \sinh(2(2n+3)\eta)}{\sinh(4\eta) \sinh(2(2n+1)\eta)} + c_{0}\,,
\end{align}
where 
\be
c_{0} = \frac{d}{du}\log\left[\phi(u,p)\, z_0(u)\, 
y_0(u,p)\right]\Big\vert_{u=0} \,.
\ee

As in the homogeneous case \cite{Nepomechie:2018nvl}, the Dynkin 
labels $[a_1^{(l)}, \ldots, a_{n-p}^{(l)}]$ of the representations of 
the ``left'' algebra $B_{n-p}$ (with $p=0, 1, \ldots, n-1$) are given by
\begin{align}
a_i^{(l)} &=m_{p+i-1}-2m_{p+i}+m_{p+i+1}\,, \qquad i=1,...,n-p-1 \,. \non \\
a_{n-p}^{(l)} &=2m_{n-1}-2m_n \,,
\label{lDynkin}
\end{align}
where $m_{0}=N$.
Similarly, the Dynkin 
labels $[a_1^{(r)}, \ldots, a_{p}^{(r)}]$ of the representations of 
the ``right'' algebra $C_{p}$ (with $p=1, 2, \ldots, n$) are given by
\begin{align}
a_i^{(r)} &=m_{i-1}-2m_i+m_{i+1} \,, \qquad i=1,\ldots ,p-1\,,  \non  \\
a_p^{(r)} &=m_{p-1}-m_p \,. \label{rDynkin}
\end{align}
Given the cardinalities of the Bethe roots of each type 
(namely, $m_{1}, \ldots, m_{n}$) for an eigenvalue $\Lambda^{(m_{1}, 
\ldots, m_{n})}(u)$, Eqs. \eqref{lDynkin}-\eqref{rDynkin} determine the Dynkin labels of the 
corresponding ``left'' and ``right'' representations, from which one 
can deduce (e.g., using LieART \cite{Feger:2019tvk}) their dimensions, and 
therefore the eigenvalue's degeneracy.

We have numerically verified the completeness of this Bethe ansatz
solution for small values of $n$ and $N$, namely, $ n=1,2 $ with $
N=1,2,3 $, and $ n=3 $ with $ N=1,2 $, for all $p=0, 1, \ldots, n$ and
$\gamma_{0}=\pm 1$, for some generic value of the anisotropy $\eta$, along the lines in \cite{Ahmed:2017mqq}.

We emphasize that in the homogeneous case \cite{Nepomechie:2018nvl}, the Q-functions are given by
$$Q^{[l]}(u)=\prod_{j=1}^{m_l}\sinh\left(\tfrac{1}{2}(u-u_j^{[l]})\right)\sinh\left(\tfrac{1}{2}(u+u_j^{[l]})\right)\,;$$
but in the presence of the boundary inhomogeneity \eqref{inhomgen}, 
the Q-functions are instead given by (\ref{Qfunc}), with a $\cosh$ 
instead of $\sinh$ in the second factor. Consequently, it is not just 
the ``left-hand-side'', but also the ``right-hand-side'' of the Bethe 
equations \eqref{BE1}-\eqref{BE3} that is affected by the boundary inhomogeneity, contrary to the 
conventional wisdom that the boundary affects only the former. 
Note also that, contrary to what usually happens for open spin chains, 
the power appearing in the left-hand-side of the first Bethe equation 
(\ref{BE1}) is $N$ instead of $2N$.

\section{Conclusions}\label{sec:end}

We have seen that an integrable open quantum spin chain with a
boundary inhomogeneity \eqref{topen} can have a local Hamiltonian
\eqref{Heven}-\eqref{Hodd}, as well as QG symmetry \eqref{QGt} that accounts for
rich degeneracies in the spectrum.  The presence of a boundary
inhomogeneity affects the crossing relation \eqref{tcrossing}, and has
a profound effect on the Bethe ansatz solution, most notably on the
Q-functions (\ref{Qfunc}).

We have focused here on a boundary inhomogeneity whose value is half
the period of the R-matrix, and with corresponding staggered bulk
inhomogeneities \eqref{inhomgen}.  It may be interesting to consider
other choices of boundary and bulk inhomogeneities, especially if they
give rise to local Hamiltonians.  Although we have also focused here
on the infinite family of models constructed with $A^{(2)}_{2n}$
R-matrices, we expect that similar results hold for other models with 
crossing symmetry, such as those considered in \cite{Nepomechie:2018dsn, Nepomechie:2018nvl}.
In the simpler $A^{(1)}_{1}$ case \cite{Nepomechie:2020onv}, the
Hamiltonian can be formulated in a beautiful compact form
\cite{Robertson:2020imc} in terms of Temperley-Lieb (TL) operators
\cite{Temperley:1971iq}.  It would be interesting if the Hamiltonians
obtained here \eqref{Heven}-\eqref{Hodd} could be reformulated in a
similar way in terms of some sort of generalized TL operators, at
least for the ``extremal'' cases $p=0, n$, where the QG symmetry is
$U_{q}(B_{n})$, $U_{q}(C_{n})$, respectively.  The continuum limit of
the $A^{(1)}_{1}$ model \cite{Nepomechie:2020onv} is described
\cite{Robertson:2020imc} by a non-compact CFT; it would be very
interesting if a similar phenomenon occurs for the higher-rank 
models introduced here.

\section*{Acknowledgments}
We thank Marius de Leeuw for his help with numerical checks of 
the Hamiltonian, and Azat Gainutdinov for helpful correspondence.
A.L.R. is supported by SFI and the EPSRC No. 18/EPSRC/3590.


\begin{thebibliography}{10}

\bibitem{Gaudin:1971zza}
M.~Gaudin, ``{Boundary energy of a Bose gas in one dimension},''
  \href{http://dx.doi.org/10.1103/PhysRevA.4.386}{{\em Phys. Rev. A} {\bfseries
  4} (1971) 386--394}.

\bibitem{Alcaraz:1987uk}
F.~C. Alcaraz, M.~N. Barber, M.~T. Batchelor, R.~J. Baxter, and G.~R.~W.
  Quispel, ``{Surface exponents of the quantum XXZ, Ashkin-Teller and Potts
  models},''
\href{http://dx.doi.org/10.1088/0305-4470/20/18/038}{{\em J. Phys.} {\bfseries
  A20} (1987) 6397}.

\bibitem{Sklyanin:1988yz}
E.~K. Sklyanin, ``{Boundary conditions for integrable quantum systems},''
\href{http://dx.doi.org/10.1088/0305-4470/21/10/015}{{\em J. Phys.} {\bfseries
  A21} (1988) 2375}.

\bibitem{Pasquier:1989kd}
V.~Pasquier and H.~Saleur, ``{Common structures between finite systems and
  conformal field theories through quantum groups},''
\href{http://dx.doi.org/10.1016/0550-3213(90)90122-T}{{\em Nucl. Phys.}
  {\bfseries B330} (1990) 523--556}.

\bibitem{Sandow:1994}
S.~Sandow, ``{Partially asymmetric exclusion process with open boundaries},''
  {\em Phys. Rev. E} {\bfseries 50} no.~4, (1994) 2660--2667,
  \href{http://arxiv.org/abs/cond-mat/9405073}{{\ttfamily
  arXiv:cond-mat/9405073 [cond-mat]}}.

\bibitem{Kitanine:2007bi}
N.~Kitanine, K.~K. Kozlowski, J.~M. Maillet, G.~Niccoli, N.~A. Slavnov, and
  V.~Terras, ``{Correlation functions of the open XXZ chain I},''
  \href{http://dx.doi.org/10.1088/1742-5468/2007/10/P10009}{{\em J. Stat.
  Mech.} {\bfseries 0710} (2007) P10009},
  \href{http://arxiv.org/abs/0707.1995}{{\ttfamily arXiv:0707.1995 [hep-th]}}.

\bibitem{Zoubos:2010kh}
K.~Zoubos, ``{Review of AdS/CFT Integrability, Chapter IV.2: Deformations,
  Orbifolds and Open Boundaries},''
  \href{http://dx.doi.org/10.1007/s11005-011-0515-8}{{\em Lett. Math. Phys.}
  {\bfseries 99} (2012) 375--400},
  \href{http://arxiv.org/abs/1012.3998}{{\ttfamily arXiv:1012.3998 [hep-th]}}.

\bibitem{Wang2015}
Y.~Wang, W.-L. Yang, J.~Cao, and K.~Shi, {\em Off-Diagonal Bethe Ansatz for
  Exactly Solvable Models}.
\newblock Springer, 2015.

\bibitem{Jimbo:1989qm}
M.~Jimbo, ``{Introduction to the {Yang-Baxter} Equation},''
  \href{http://dx.doi.org/10.1142/S0217751X89001503}{{\em Int. J. Mod. Phys. A}
  {\bfseries 4} (1989) 3759--3777}.

\bibitem{Cherednik:1985vs}
I.~V. Cherednik, ``{Factorizing particles on a half line and root systems},''
  \href{http://dx.doi.org/10.1007/BF01038545}{{\em Theor. Math. Phys.}
  {\bfseries 61} (1984) 977--983}.
[Teor. Mat. Fiz.61,35 (1984)].

\bibitem{Ghoshal:1993tm}
S.~Ghoshal and A.~B. Zamolodchikov, ``{Boundary S matrix and boundary state in
  two-dimensional integrable quantum field theory},''
  \href{http://dx.doi.org/10.1142/S0217751X94001552}{{\em Int. J. Mod. Phys.}
  {\bfseries A9} (1994) 3841--3886},
  \href{http://arxiv.org/abs/hep-th/9306002}{{\ttfamily arXiv:hep-th/9306002
  [hep-th]}}.
[Erratum: Int. J. Mod. Phys.A9,4353 (1994)].

\bibitem{Nepomechie:2020onv}
R.~I. Nepomechie and A.~L. Retore, ``{Factorization identities and algebraic
  Bethe ansatz for $ {D}_2^{(2)} $ models},''
  \href{http://dx.doi.org/10.1007/JHEP03(2021)089}{{\em JHEP} {\bfseries 03}
  (2021) 089}, \href{http://arxiv.org/abs/2012.08367}{{\ttfamily
  arXiv:2012.08367 [hep-th]}}.

\bibitem{Chari:1994pz}
V.~Chari and A.~Pressley, {\em {A guide to quantum groups}}.
\newblock Cambridge University Press,
1994.
\newblock

\bibitem{Robertson:2020imc}
N.~F. Robertson, J.~L. Jacobsen, and H.~Saleur, ``{Lattice regularisation of a
  non-compact boundary conformal field theory},''
  \href{http://dx.doi.org/10.1007/JHEP02(2021)180}{{\em JHEP} {\bfseries 02}
  (2021) 180}, \href{http://arxiv.org/abs/2012.07757}{{\ttfamily
  arXiv:2012.07757 [hep-th]}}.

\bibitem{Jacobsen:2005xz}
J.~L. Jacobsen and H.~Saleur, ``{The antiferromagnetic transition for the
  square-lattice Potts model},''
  \href{http://dx.doi.org/10.1016/j.nuclphysb.2006.02.041}{{\em Nucl. Phys. B}
  {\bfseries 743} (2006) 207--248},
  \href{http://arxiv.org/abs/cond-mat/0512058}{{\ttfamily
  arXiv:cond-mat/0512058}}.

\bibitem{Ikhlef:2008zz}
Y.~Ikhlef, J.~Jacobsen, and H.~Saleur, ``{A staggered six-vertex model with
  non-compact continuum limit},''
  \href{http://dx.doi.org/10.1016/j.nuclphysb.2007.07.004}{{\em Nucl. Phys. B}
  {\bfseries 789} (2008) 483--524},
  \href{http://arxiv.org/abs/cond-mat/0612037}{{\ttfamily
  arXiv:cond-mat/0612037 [cond-mat]}}.

\bibitem{Ikhlef:2011ay}
Y.~Ikhlef, J.~L. Jacobsen, and H.~Saleur, ``{An Integrable spin chain for the
  SL(2,R)/U(1) black hole sigma model},''
  \href{http://dx.doi.org/10.1103/PhysRevLett.108.081601}{{\em Phys. Rev.
  Lett.} {\bfseries 108} (2012) 081601},
  \href{http://arxiv.org/abs/1109.1119}{{\ttfamily arXiv:1109.1119 [hep-th]}}.

\bibitem{Candu:2013fva}
C.~Candu and Y.~Ikhlef, ``{Nonlinear integral equations for the SL(2,
  $\mathbb{R})$/U(1) black hole sigma model},''
  \href{http://dx.doi.org/10.1088/1751-8113/46/41/415401}{{\em J. Phys. A}
  {\bfseries 46} (2013) 415401},
  \href{http://arxiv.org/abs/1306.2646}{{\ttfamily arXiv:1306.2646 [hep-th]}}.

\bibitem{Frahm:2013cma}
H.~Frahm and A.~Seel, ``{The Staggered Six-Vertex Model: Conformal Invariance
  and Corrections to Scaling},''
  \href{http://dx.doi.org/10.1016/j.nuclphysb.2013.12.015}{{\em Nucl. Phys. B}
  {\bfseries 879} (2014) 382--406},
  \href{http://arxiv.org/abs/1311.6911}{{\ttfamily arXiv:1311.6911
  [cond-mat.stat-mech]}}.

\bibitem{Bazhanov:2019xvy}
V.~V. Bazhanov, G.~A. Kotousov, S.~M. Koval, and S.~L. Lukyanov, ``{On the
  scaling behaviour of the alternating spin chain},''
  \href{http://dx.doi.org/10.1007/JHEP08(2019)087}{{\em JHEP} {\bfseries 08}
  (2019) 087}, \href{http://arxiv.org/abs/1903.05033}{{\ttfamily
  arXiv:1903.05033 [hep-th]}}.

\bibitem{Bazhanov:2020dlm}
V.~V. Bazhanov, G.~A. Kotousov, S.~M. Koval, and S.~L. Lukyanov, ``{Scaling
  limit of the ${\cal Z}_2$ invariant inhomogeneous six-vertex model},''
  \href{http://arxiv.org/abs/2010.10613}{{\ttfamily arXiv:2010.10613
  [math-ph]}}.

\bibitem{Bazhanov:1984gu}
V.~V. Bazhanov, ``{Trigonometric solution of triangle equations and classical
  Lie algebras},''
\href{http://dx.doi.org/10.1016/0370-2693(85)90259-X}{{\em Phys. Lett.}
  {\bfseries B159} (1985) 321--324}.

\bibitem{Bazhanov:1986mu}
V.~V. Bazhanov, ``{Integrable quantum systems and classical Lie algebras},''
\href{http://dx.doi.org/10.1007/BF01221256}{{\em Commun. Math. Phys.}
  {\bfseries 113} (1987) 471--503}.

\bibitem{Jimbo:1985ua}
M.~Jimbo, ``{Quantum R matrix for the generalized Toda system},''
\href{http://dx.doi.org/10.1007/BF01221646}{{\em Commun. Math. Phys.}
  {\bfseries 102} (1986) 537--547}.

\bibitem{Nepomechie:2018dsn}
R.~I. Nepomechie and A.~L. Retore, ``{Surveying the quantum group symmetries of
  integrable open spin chains},''
  \href{http://dx.doi.org/10.1016/j.nuclphysb.2018.02.023}{{\em Nucl. Phys.}
  {\bfseries B930} (2018) 91--134},
\href{http://arxiv.org/abs/1802.04864}{{\ttfamily arXiv:1802.04864 [hep-th]}}.

\bibitem{Nepomechie:2018nvl}
R.~I. Nepomechie and A.~L. Retore, ``{The spectrum of quantum-group-invariant
  transfer matrices},''
  \href{http://dx.doi.org/10.1016/j.nuclphysb.2018.11.017}{{\em Nucl. Phys.}
  {\bfseries B938} (2019) 266--297},
\href{http://arxiv.org/abs/1810.09048}{{\ttfamily arXiv:1810.09048 [hep-th]}}.

\bibitem{Izergin:1980pe}
A.~G. Izergin and V.~E. Korepin, ``{The inverse scattering method approach to
  the quantum Shabat-Mikhailov model},''
\href{http://dx.doi.org/10.1007/BF01208496}{{\em Commun. Math. Phys.}
  {\bfseries 79} (1981) 303}.

\bibitem{Batchelor:1996np}
M.~T. Batchelor, V.~Fridkin, A.~Kuniba, and Y.~K. Zhou, ``{Solutions of the
  reflection equation for face and vertex models associated with $A^{(1)}_n$,
  $B^{(1)}_n$, $C^{(1)}_n$, $D^{(1)}_n$ and $A^{(2)}_n$},''
  \href{http://dx.doi.org/10.1016/0370-2693(96)00319-X}{{\em Phys. Lett.}
  {\bfseries B376} (1996) 266--274},
\href{http://arxiv.org/abs/hep-th/9601051}{{\ttfamily arXiv:hep-th/9601051
  [hep-th]}}.

\bibitem{LimaSantos:2002ui}
A.~Lima-Santos, ``{$B_{n}^{(1)}$ and $A_{2n}^{(2)}$ reflection K matrices},''
  \href{http://dx.doi.org/10.1016/S0550-3213(03)00042-7}{{\em Nucl. Phys.}
  {\bfseries B654} (2003) 466--480},
\href{http://arxiv.org/abs/nlin/0210046}{{\ttfamily arXiv:nlin/0210046
  [nlin-si]}}.

\bibitem{Malara:2004bi}
R.~Malara and A.~Lima-Santos, ``{On $A_{n-1}^{(1)}$, $B_{n}^{(1)}$,
  $C_{n}^{(1)}$, $D_{n}^{(1)}$, $A_{2n}^{(2)}$, $A_{2n-1}^{(2)}$ and
  $D_{n+1}^{(2)}$ reflection K-matrices},''
  \href{http://dx.doi.org/10.1088/1742-5468/2006/09/P09013}{{\em J. Stat.
  Mech.} {\bfseries 0609} (2006) P09013},
\href{http://arxiv.org/abs/nlin/0412058}{{\ttfamily arXiv:nlin/0412058
  [nlin-si]}}.

\bibitem{Mezincescu:1990ui}
L.~Mezincescu and R.~I. Nepomechie, ``{Integrability of open spin chains with
  quantum algebra symmetry},''
  \href{http://dx.doi.org/10.1142/S0217751X91002458,
  10.1142/S0217751X9200257X}{{\em Int. J. Mod. Phys.} {\bfseries A6} (1991)
  5231--5248}, \href{http://arxiv.org/abs/hep-th/9206047}{{\ttfamily
  arXiv:hep-th/9206047 [hep-th]}}.
[Addendum: {\em Int. J. Mod. Phys.} {\bf A7} (1992) 5657].

\bibitem{Ahmed:2017mqq}
I.~Ahmed, R.~I. Nepomechie, and C.~Wang, ``{Quantum group symmetries and
  completeness for $A_{2n}^{(2)}$ open spin chains},''
  \href{http://dx.doi.org/10.1088/1751-8121/aa7606}{{\em J. Phys.} {\bfseries
  A50} no.~28, (2017) 284002},
\href{http://arxiv.org/abs/1702.01482}{{\ttfamily arXiv:1702.01482 [math-ph]}}.

\bibitem{Baxter:1982b}
R.~J. Baxter, ``{Critical antiferromagnetic square-lattice Potts model},'' {\em
  Proc. R. Soc. Lond. A.} {\bfseries 383} no.~1784, (1982) 43--54.

\bibitem{Gainutdinov:2021}
A.~M. Gainutdinov, R.~I. Nepomechie, and A.~L. Retore, ``{Quasi quantum group symmetry
in integrable spin chains"}, in preparation. 

\bibitem{Reshetikhin:1987}
N.~{\relax Yu}. Reshetikhin, ``{The spectrum of the transfer matrices connected
  with Kac-Moody algebras},'' {\em Lett. Math. Phys.} {\bfseries 14} (1987)
  235.

\bibitem{Feger:2019tvk}
R.~Feger, T.~W. Kephart, and R.~J. Saskowski, ``{LieART 2.0 \textendash{} A
  Mathematica application for Lie Algebras and Representation Theory},''
  \href{http://dx.doi.org/10.1016/j.cpc.2020.107490}{{\em Comput. Phys.
  Commun.} {\bfseries 257} (2020) 107490},
  \href{http://arxiv.org/abs/1912.10969}{{\ttfamily arXiv:1912.10969
  [hep-th]}}.

\bibitem{Temperley:1971iq}
H.~Temperley and E.~Lieb, ``{Relations between the 'percolation' and
  'colouring' problem and other graph-theoretical problems associated with
  regular planar lattices: some exact results for the 'percolation' problem},''
  \href{http://dx.doi.org/10.1098/rspa.1971.0067}{{\em Proc. Roy. Soc. Lond. A}
  {\bfseries A322} (1971) 251--280}.

\end{thebibliography}
\providecommand{\href}[2]{#2}\begingroup\raggedright\endgroup

\end{document}